\documentclass[a4paper, 10pt, 3p]{elsarticle}
\makeatletter
\def\ps@pprintTitle{%
 \let\@oddhead\@empty{}
 \let\@evenhead\@empty{}
 \def\@oddfoot{}%
 \let\@evenfoot\@oddfoot}
\makeatother

\usepackage[utf8]{inputenc}
\usepackage{xcolor}
\usepackage{amsmath}
\usepackage{amsfonts}
\usepackage{amsthm}
\usepackage{siunitx}
\usepackage{subcaption}
\usepackage{mdframed}
\usepackage{natbib}
\usepackage{booktabs}
\usepackage{graphicx}
\usepackage{enumitem}
\usepackage{hyperref}

\newcommand{\figref}[1]{Figure~\ref{#1}}

\newcommand{\tbref}[1]{Table~\ref{#1}}

\DeclareMathOperator{\tr}{tr}

\newcommand\bs[1]{\boldsymbol{#1}}

\newcommand\stress{\boldsymbol{\sigma}}

\newcommand\strain{\bf{E}}

\newcommand\Gc{{\mathcal{G}_c}}

\newcommand\set[1]{#1}

\newcommand\body{\set{\Omega}}

\theoremstyle{remark}
\newtheorem*{remark}{Remark}

\newcommand{\coh}{\texttt{Coh2019}}
\newcommand{\nuc}{\texttt{NucCe2020}}
\newcommand{\latest}{\texttt{NucCe2024}} 
\usepackage{pgfplots}
\usetikzlibrary{spy} 
\usetikzlibrary{shapes.misc}

\usepgfplotslibrary{colorbrewer}
\usepgfplotslibrary{colormaps}
\usepgfplotslibrary{fillbetween}
\usepgfplotslibrary{groupplots}
\usetikzlibrary{patterns}
\usetikzlibrary{spy}
\usetikzlibrary{calc}
\usetikzlibrary{shapes.geometric, arrows}
\usetikzlibrary{matrix}
\usetikzlibrary{arrows,patterns}
\usepackage{environ}
\makeatletter
\newsavebox{\measure@tikzpicture}
\NewEnviron{scaletikzpicturetowidth}[1]{%
    \def\tikz@width{#1}%
    \begin{lrbox}{\measure@tikzpicture}%
        \BODY
    \end{lrbox}%
    \pgfmathparse{#1/\wd\measure@tikzpicture}%
    \BODY
}
\makeatother
\pgfplotsset{select coords between index/.style 2 args={
            x filter/.code={
                    \ifnum\coordindex<#1\fi
                    \ifnum\coordindex>#2\fi
                }
        }}
\pgfplotsset{
    colormap name=jet
}
\pgfplotsset{compat=newest}

\usepackage{tikz}

\usetikzlibrary{external}
\usetikzlibrary{matrix}
\pgfplotsset{compat=newest}

\tikzstyle{load}   = [ultra thick,-latex]
\tikzstyle{stress} = [-latex]
\tikzstyle{dim}    = [latex-latex]
\tikzstyle{axis}   = [-latex,black!55]

\tikzstyle{isometric}=[x={(0.710cm,-0.410cm)},y={(-0.710cm,-0.410cm)},z={(0cm,0.820cm)}]
\tikzstyle{dimetric} =[x={(0.935cm,-0.118cm)},y={(0cm,0.943cm)},z={(-0.354cm,-0.312cm)}]
\tikzstyle{dimetric2}=[x={(0.935cm,-0.118cm)},z={(0cm,0.943cm)},y={(+0.354cm,+0.312cm)}]
\tikzstyle{trimetric}=[x={(0.926cm,-0.207cm)},y={(0cm,0.837cm)},z={(-0.378cm,-0.507cm)}]

\tikzstyle{startstop} = [rectangle, rounded corners, minimum width=3cm, minimum height=1cm,text centered, draw=black, fill=red!30]
\tikzstyle{io} = [trapezium, trapezium left angle=70, trapezium right angle=110, minimum width=3cm, minimum height=1cm, text centered, draw=black, fill=blue!30]
\tikzstyle{process} = [rectangle, minimum width=3cm, minimum height=1cm, text centered, draw=black, fill=orange!30]
\tikzstyle{decision} = [diamond, minimum width=3cm, minimum height=1cm, text centered, draw=black, fill=green!30]
\tikzstyle{arrow} = [thick,->,>=stealth] 
\graphicspath{{./posts/figures}}

\begin{document}
\begin{frontmatter}


\title{On the effects of material strength in dynamic fracture: A phase-field study}

\author{Yangyuanchen Liu} 
\ead{yangyc.liu@duke.edu}
    
\author{Oscar Lopez-Pamies} 
\ead{pamies@illinois.edu}
  
\author{John E.\ Dolbow}
\ead{jdolbow@duke.edu}

\address{Department of Mechanical Engineering, Duke University, Durham, NC 27708, USA \vspace{0.1cm}}

\address{Department of Civil and Environmental Engineering, University of Illinois, Urbana--Champaign, IL 61801, USA  \vspace{0.1cm}}

\begin{abstract}


Over the past seven years, full-field analyses of a wide range of classical as well as modern quasi-static fracture experiments on nominally elastic brittle materials --- ranging from hard ceramics to soft elastomers --- have repeatedly identified the material strength surface as one of the key material properties that governs not only the nucleation of cracks, but also their propagation. Central to these analyses are the results generated by the Griffith phase-field fracture theory with material strength introduced in \cite{KFLP18,kumar2020elastomers,kumar2020}. The first of two objectives of this paper is to extend this theory to account for inertia, this for the basic case of isotropic linear elastic brittle materials. From an applications point of view, the theory amounts to solving an initial-boundary-value problem comprised of a  hyperbolic PDE coupled with an elliptic PDE for the displacement
field $\mathbf{u}(\mathbf{X},t)$ and the phase field $d(\mathbf{X},t)$. A robust scheme is presented to generate solutions for these equations that is based on an adaptive finite-element discretization of space and an implicit finite-difference discretization of time. 
The second objective is to illustrate the descriptive and predictive capabilities of the proposed theory via simulations of benchmark problems and experiments. These include problems involving fracture nucleation from large pre-existing cracks, such as the classical Kalthoff-Winkler experiments, as well as problems involving fracture nucleation within the bulk, such as the dynamic Brazilian fracture experiments. As expected from the analyses of quasi-static experiments, the simulations make it clear that accounting for the material strength surface as an independent macroscopic property remains essential to explaining and describing experimental observations when inertia is not negligible. This key insight is further illustrated by means of comparisons with results from  classical phase-field models, which, by construction, are unable to account for a strength surface that is independent of the elasticity and toughness of the material.

\end{abstract}

\end{frontmatter}

\vspace{-0.1cm}

\section{Introduction}

In a string of recent contributions, Kumar, Francfort, and Lopez-Pamies~\cite{KFLP18}, Kumar and Lopez-Pamies~\cite{kumar2020elastomers}, and Kumar et al.~\cite{kumar2020} established that any potentially successful attempt at a macroscopic theoretical description of nucleation and propagation of fracture in elastic brittle
materials must be able to account for any elasticity, any strength, and any toughness that the material of interest may exhibit. For the basic case of interest here, that of isotropic linear elastic brittle materials, this means that any potentially successful theory of fracture must account for any stored-energy function
\begin{equation}
\label{eq:storednrg}
W({\bf{E}}) = \mu\, \mbox{tr}{\bf{E}}^2 + \frac{\lambda}{2} \left( \mbox{tr} {\bf{E}} \right)^2
\end{equation}
describing the elasticity of the material, any strength surface
\begin{equation}
\label{eq:strsurf}
\mathcal{F} \left( \sigma_1, \sigma_2, \sigma_3 \right ) = 0
\end{equation}
describing the set of critical stresses at which the material fractures under states of spatially uniform stress,\footnote{In these expressions, making use of standard notation, ${\bf{E}}$ stands for the infinitesimal strain tensor, $\mu$ and $\lambda$ denote the Lam\'e constants, and $\sigma_1, \sigma_2, \sigma_3$ stand for the principal Cauchy stresses.}  and any fracture toughness
\begin{equation}
\label{eq:critnrg}
\Gc
\end{equation}
describing the intrinsic resistance for crack growth in the material. Guided by this insight, these same works introduced a phase-field theory for the nucleation and propagation of fracture in elastic brittle materials under quasi-static loading conditions. Over the past seven years, numerous studies  have provided an extensive set of validation results across various nominally elastic brittle materials, geometries, and loading scenarios. These span studies of crack nucleation in soft materials such as silicones \cite{KFLP18,kumar2018} and synthetic rubbers \cite{kamarei2024}, to studies of crack nucleation in hard materials such as graphite \cite{kumar2020} and glass \cite{kumar2022}. The theory has also been used to explain the poker-chip experiments of Gent and Lindley \cite{kumar2021poker} on natural rubber and to develop a new protocol to deduce the tensile strength of a material from a Brazilian test \cite{kumar2024brz}.  Collectively, these works suggest that the theory is complete in that it is able to describe and predict where and when fracture nucleates and propagates in elastic brittle materials at large under general quasi-static conditions. The studies have also served to repeatedly identify the material strength surface as one of the key material properties that governs not only the nucleation of cracks, but also their propagation. 


In this context, the first objective of this paper is to extend the phase-field fracture theory introduced in \cite{KFLP18,kumar2020elastomers,kumar2020} to account for inertia. We do so in Section \ref{sec:form} for the basic case of isotropic linear elastic brittle materials. As shown there, in terms of material properties, in addition to the stored-energy function (\ref{eq:storednrg}), the strength surface (\ref{eq:strsurf}), and the fracture toughness (\ref{eq:critnrg}) of the material, one must also simply account for its mass density 
\begin{equation}
\label{eq:density}
    \rho.
\end{equation}
The theory is given in the form of an initial-boundary-value problem comprised of a  hyperbolic PDE coupled with an elliptic PDE for the displacement
field $\mathbf{u}(\mathbf{X},t)$ and the phase field $d(\mathbf{X},t)$. In Section \ref{sec:fem}, we outline a robust scheme to generate solutions for these equations that is based on an adaptive finite-element discretization of space and an implicit finite-difference discretization of time. 

The second objective of this paper is to illustrate the descriptive and predictive capabilities of the proposed theory via simulations of benchmark problems and experiments. We do so in Sections \ref{sec:numer} and \ref{sec:experiments}. Where appropriate, comparisons are made with classical phase-field models of dynamic fracture employing various energy splits. Emphasis is placed on the effective strength surface encapsulated within the various models and on how this impacts the accompanying simulations. 

\subsection{A comment on existing related approaches for dynamic fracture in elastic brittle materials}

Before proceeding with the presentation of the proposed phase-field theory, it is instructive to summarize the various phase-field models that have been put forth in the literature for dynamic fracture in elastic brittle materials. Invariably, all such phase-field models correspond to extensions of classical variational phase-field models\footnote{By classical phase-field models we mean phase-field models of fracture that $\Gamma$-converge to the variational theory of brittle fracture of Francfort and Marigo~\cite{francfort1998}.} in which inertia is simply added to the balance of linear momentum. Consequently, as recently reviewed in \cite{Pamies2024nucleation,JAMNote}, all such phase-field models are fundamentally incomplete since they cannot account for material strength as an independent material property. 

The first phase-field model for dynamic fracture appears to be that proposed by Bourdin et al.~\cite{larsen2011dynamic}; see also Larsen et al.~\cite{larsen2010dynamic}. This model, being a direct extension of the classical \texttt{AT}$_2$ variational phase-field model \cite{bourdin2000numerical}, does not employ any split of the elastic energy. As a result, it is unable to prevent crack growth into regions of large compressive strains.

Soon thereafter, the \texttt{AT}$_2$ phase-field model by Miehe et al.~\cite{miehe2010rate} employing a spectral split was extended by Hofacker and Miehe \cite{hofacker2012,hofacker2013} to dynamic fracture. Around the same time, Borden et al.~\cite{borden2012dynamic} proposed an alternative \texttt{AT}$_2$ phase-field model based on a different energy split. More recently, a  number of works \cite{bleyer_2017,geelen2019,WuDynamic2020} have also examined \texttt{AT}$_1$ phase-field models and cohesive-type degradation functions \cite{lorentz2011convergence} with a variety of energy splits. Like any phase-field model that uses energy splits, all of these models improve on the corresponding base models without energy splits in that they may prevent compressive cracks under some subset of loading conditions, but not in general \cite{Pamies2024nucleation}.

Beyond phase-field models of fracture, the present work falls into the broad space of models and numerical methods that have been advanced over the past several decades on fracture, with a focus on dynamic loading conditions in particular. These include, but are not limited to, existing versions of cohesive fracture formulations~\cite{camacho1996,WuDynamic2020}, sharp and regularized the same, peridynamic models~\cite{silling2000}, and eigenfracture approaches~\cite{schmidt2009}.  Although all of these models have been adopted \emph{mutatis mutandis} for applications in dynamic fracture, much like the aforementioned classical phase-field models, they do not seem capable of accounting for arbitrary strength surfaces. This may be due to their original intent being focused on fracture propagation and not on fracture nucleation and, within the conventional wisdom of the latter, to the misunderstanding of strength as an intrinsic macroscopic material property.  In this work, as an example, we contrast results obtained using the proposed phase-field model to those described by Mehrmashhadi et al.~\cite{mehrmashhadi_soda_peridynamic_2020} using peridynamics and a classical phase-field model, for simulating the dynamic fracture of soda-lime glass specimens.  Given their basis in a Griffith model of fracture, at least for crack propagation, there is no reason to expect these models to yield qualitatively different results, and indeed, contrary to the claims stated by Mehrmashhadi et al.~\cite{mehrmashhadi_soda_peridynamic_2020}, this is what our results show.  

\section{The proposed phase-field model for dynamic fracture in elastic brittle materials}
\label{sec:form}

\begin{figure}[htbp!]
  \centering
  \includegraphics[width=.9\columnwidth]{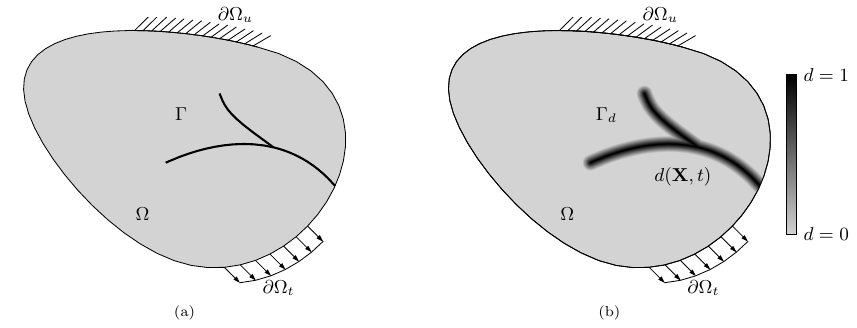}
  \caption{Schematic representation of a body $\Omega$, in its undeformed stress-free configuration at time $t=0$, with (a) a pre-existing sharp crack $\Gamma$, and (b) its regularized representation via the phase field $d$.}
  \label{fig:pf_scheme}
\end{figure}


%
\subsection{General considerations}

Consider a body made of an isotropic linear elastic brittle material, with stored-energy function \eqref{eq:storednrg}, strength surface \eqref{eq:strsurf}, critical energy release rate \eqref{eq:critnrg}, and mass density \eqref{eq:density} that occupies an open bounded domain $\Omega \subset \mathbb{R}^n$ ($n\in \{1,2,3\}$), with boundary $\partial\Omega$ and unit outward normal ${\bf{N}}$, in its undeformed and stress-free configuration at time $t=0$; see  \figref{fig:pf_scheme}.  At a subsequent time $t\in(0,T]$, due to an externally applied displacement $\bar{\bf{u}}({\bf{X}},t)$ on a part $\partial \Omega_u$ of the boundary and a traction $\bar{\bf{t}}({\bf{X}},t)$ on the complementary part $\partial \Omega_t = \partial \Omega \backslash \partial\Omega_u$, the position vector ${\bf{X}}$ of a material point moves to a new position specified by 
\begin{equation}
    {\bf{x}} = {\bf{X}} + {\bf{u}}({\bf{X}},t )
\end{equation}
in terms of the displacement field ${\bf{u}}$.  We assume small deformations and deformation gradients, and write the infinitesimal strain tensor as 
\begin{equation}
    {\bf{E}}({\bf{u}}) = \frac{1}{2} \left( \nabla {\bf{u}} + \nabla {\bf{u}}^T   \right).
\end{equation}

In response to the same externally applied loads that give rise to the aforementioned deformation, cracks can also nucleate and propagate in the body. These cracks are regularized by an order parameter or phase field 
\begin{equation}
    d = d({\bf{X}},t),
\end{equation}
taking values in $[0,1]$.  Regions with $d=0$ correspond to fully intact material, fully fractured regions are represented by $d=1$, while the transition from $d=0$ to $d=1$ is set to occur smoothly over regions of small thickness of regularization length scale
$\ell$; see \figref{fig:pf_scheme}.  


\subsection{The governing equations}

The governing equations are a straightforward extension of the theory put forth in \cite{kamarei2024,larsen2024variational}: one simply needs to account for the presence of inertial effects in the balance of linear momentum. 
Accordingly, the displacement field ${\bf{u}}({\bf{X}},t)$ and phase field $d({\bf{X}},t)$ are determined by the coupled system of partial differential equations (PDEs) given by the linear momentum balance
\begin{equation}
  \label{eq:mombal}
  \mbox{Div} \left[  (1-d)^2 \frac{\partial W}{\partial {\bf{E}}} \left( {\bf{E}} ({\bf{u}}) \right)  \right] =   \rho  \ddot{\bf{u}}, \quad ({\bf{X}},t)\in\Omega\times[0,T],
\end{equation}
and the phase-field evolution equation
\begin{equation}
  \left\{
  \begin{array}{ll}
    \ell \delta^{\ell} \Gc \Delta d = -\dfrac{8}{3}(1-d) W\left( {\bf{E}} ({\bf{u}}) \right) + \dfrac{4}{3} c_e ({\bf{X}},t)  + \dfrac{\delta^\ell\Gc}{2\ell}  & \quad\textrm{if}\quad  \dot{d}({\bf{X}}, t) > 0 \vspace{0.2cm}\\
     \ell \delta^{\ell} \Gc \Delta d \le -\dfrac{8}{3}(1-d) W\left( {\bf{E}} ({\bf{u}}) \right)  + \dfrac{4}{3} c_e ({\bf{X}},t) + \dfrac{\delta^\ell\Gc}{2\ell}  & \quad\textrm{if}\quad \dot{d}({\bf{X}}, t) = 0
  \vspace{0.2cm}\\ \dot{d} ({\bf{X}},t)  = 0 & \quad\textrm{if}  \quad d({\bf{X}},t) =1 \end{array}\right., \quad ({\bf{X}},t)\in\Omega\times[0,T],
  \label{eq:damage_evolution}
\end{equation}
supplemented by boundary and initial conditions, as detailed below.  In \eqref{eq:damage_evolution}, $\ell > 0$ is the regularization or localization length, $c_e ({\bf{X}},t)$ is a driving force whose specific constitutive prescription depends on the particular form of the strength surface \eqref{eq:strsurf}, and $\delta^{\ell} > 0$ is a coefficient whose specific constitutive prescription depends in turn on the particular form of $c_e ({\bf{X}},t)$.  

The above PDEs are supplemented by the boundary conditions
\begin{equation}
\label{eq:bcs}
    \left\{ \begin{array}{llc} {\bf{u}} = {\bar{\bf{u}}}({\bf{X}},t), &  ({\bf{X}},t)\in\partial\Omega_u \times [0,T]  \\[8pt] \left[  (1-d)^2 \dfrac{\partial W}{\partial {\bf{E}}} \left( {\bf{E}} ({\bf{u}}) \right)  \right] {\bf{N}} = {\bar{\bf{t}}} ({\bf{X}},t),  & ({\bf{X}},t)\in\partial\Omega_t \times [0,T] \\[8pt] \nabla d \cdot {\bf{N}} = 0, &  ({\bf{X}},t)\in\partial\Omega \times [0,T]  \end{array} \right., 
\end{equation}
and the initial conditions
\begin{equation}
\label{eq:ics}
\left\{ \begin{array}{llc} {\bf{u}}({\bf{X}},0)  = {\bf{u}}_0({\bf{X}}),  & {\bf{X}}\in\Omega  \\[8pt] 
\dot{{\bf{u}}}({\bf{X}},0)  = {\bf{v}}_0 ({\bf{X}}), & {\bf{X}}\in\Omega  \\[8pt]  d ({\bf{X}},0)  = d_0 ({\bf{X}}), & {\bf{X}}\in\Omega   \end{array} \right.,
\end{equation}
for both the displacement and phase field.  

\begin{remark}
    The localization length $\ell$ in \eqref{eq:damage_evolution} is a regularization parameter that is devoid of any physical meaning.  In practice, it is selected to be significantly smaller than the smallest characteristic length scale in the structural problem at hand, as well as no significantly larger than the smallest material length scale built in to \eqref{eq:mombal}-\eqref{eq:damage_evolution}.  Provided that $\ell$ is selected in this manner, the predictions generated by discretizations of \eqref{eq:mombal}-\eqref{eq:ics} are essentially independent of $\ell$; see, e.g., the Appendix in \cite{kumar2022}.  This insensitivity for sufficiently small $\ell$ is also examined in the numerical results section of this manuscript.  
\end{remark}

\begin{remark}
The inequalities in \eqref{eq:damage_evolution} embody the classical assumption that fracture is a purely dissipative and irreversible process.  
\end{remark}

\begin{remark}
The PDEs (\ref{eq:mombal})-(\ref{eq:damage_evolution}) do \textit{not} employ any split of the stored-energy function $W(\bf{E})$. The growth of cracks into regions of large compressive strain is prevented by accounting for the strength surface (\ref{eq:strsurf}) via the driving force $c_e$.
\end{remark}


\begin{figure}[b!]
  \centering
  \includegraphics[width=\columnwidth]{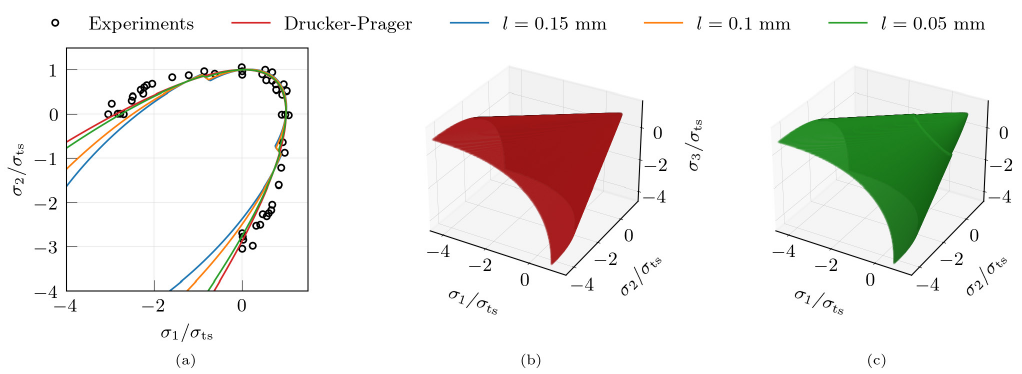}
  \caption{Comparisons between the experimental strength data for graphite by Sato et al.~\cite{sato1987graphite}, the Drucker-Prager strength surface (\ref{eq:DP}) fitted to that data, 
    and the corresponding strength surface (\ref{eq:strength}) predicted by the  phase-field theory (\ref{eq:mombal})-(\ref{eq:ics}) with driving force (\ref{eq:ce}) and coefficient (\ref{eq:delta}).
    (a) Strength surfaces in plane-stress states with $\sigma_3 = 0$. (b) The Drucker-Prager strength surface (\ref{eq:DP}) and
    (c) the strength surface (\ref{eq:strength}) predicted by the  phase-field theory (\ref{eq:mombal})-(\ref{eq:ics}) in the principal stress space ($\sigma_1, \sigma_2, \sigma_3$).
    The results are plotted for three decreasing values of the regularization length $\ell$.
  }
  \label{fig:strength}
\end{figure}

\subsection{The driving force $c_e$ and the coefficient $\delta^{\ell}$ } 

For a solid whose strength surface \eqref{eq:strsurf} is characterized by a Drucker-Prager~\cite{drucker1952} strength surface
\begin{equation}
\label{eq:DP}
\mathcal{F} \left( \sigma_1, \sigma_2, \sigma_3 \right ) = \sqrt{\dfrac{1}{3}(\sigma_1^2+\sigma_2^2+\sigma_3^2-\sigma_1\sigma_2-\sigma_1\sigma_3-\sigma_2\sigma_3)}+\gamma_1(\sigma_1+\sigma_2+\sigma_3)+\gamma_0=0,
\end{equation}
where $\gamma_1$ and $\gamma_0$ are materials constants, Kumar et al.~\cite{kumar2020,kumar2022} and Kamarei et al.~\cite{kamarei2024} have worked out various constitutive prescriptions for the driving force $c_e({\bf{X}},t)$ and the coefficient $\delta^{\ell}$ that are equivalent in the limit as $\ell \searrow 0$, with different corrections of $O(\ell^0)$. Here, we make use of the constitutive prescription proposed by Kamarei et al.~\cite{kamarei2024}.  It reads
\begin{equation}
\label{eq:ce}
  c_e ({\bf{X}},t) = \widehat{c}_e(\mathcal{I}_1, \mathcal{J}_2,d;\ell)=\alpha_2 \sqrt{\mathcal{J}_2} +  \alpha_1 \mathcal{I}_1 + (1-d)\left(1 - \dfrac{\sqrt{\mathcal{I}_1^2}}{\mathcal{I}_1}\right)W({\bf{E}}({\bf{u}}))
\end{equation}
with
\begin{equation}
\label{eq:delta}
\delta^\ell = \left(\frac{\sigma_{\text{ts}}+(1+2\sqrt{3})\sigma_{\text{hs}}}{(8+3\sqrt{3})\sigma_{\text{hs}}}\right)\frac{3\Gc}{16 \mathcal{W}_{\text{ts}} \ell} + \frac{2}{5},
\end{equation}
where 
\begin{equation}
  \left\{\begin{array}{l}
      \alpha_1 = -\dfrac{1}{\sigma_{\text{hs}}}\delta^l \dfrac{\Gc}{8\ell} + \dfrac{2\mathcal{W}_{\text{hs}}}{3\sigma_{\text{hs}}} \vspace{0.2cm}\\
      \alpha_2 = -\dfrac{\sqrt{3}(3\sigma_{\text{hs}} - \sigma_{\text{ts}})}{\sigma_{\text{hs}}\sigma_{\text{ts}}}\delta^\ell \dfrac{\Gc}{8l} - \dfrac{2\mathcal{W}_{\text{hs}}}{\sqrt{3}\sigma_{\text{hs}}} + \dfrac{2\sqrt{3}\mathcal{W}_{\text{ts}}}{\sigma_{\text{ts}}} 
        \end{array}\right.,\quad\left\{\begin{array}{l}
      \mathcal{W}_{\text{ts}} = \dfrac{\sigma_{\text{ts}}^2}{2E} \vspace{0.2cm}\\ \mathcal{W}_{\text{hs}} = \dfrac{\sigma_{\text{hs}}^2}{2\kappa}\end{array}\right.,
\end{equation}
and
\begin{equation}
    \left\{ \begin{array}{l} \mathcal{I}_1 = \tr \stress = 3 \kappa (1-d)^2 \tr {\bf{E}} \\[8pt] \mathcal{J}_2 = \frac{1}{2} \tr \stress_D^2 = 2\mu^2 (1-d)^4 \tr {\bf{E}}_D^2, \quad  {\bf{E}}_D = {\bf{E}} - \frac{1}{3} \left( \tr {\bf{E}} \right) {\bf{I}} \end{array} \right. ,
\end{equation}
and where we have made use of the classical relationships $E = \mu(3\lambda + 2\mu)/(\lambda + \mu)$ and $\kappa = \lambda + \frac{2}{3} \mu $ between Young's modulus $E$ and bulk modulus $\kappa$ and the Lam\'e constants $\mu$ and $\lambda$. For definiteness, the above prescription has made use of the uniaxial tensile strength $\sigma_{\text{ts}}$ and the hydrostatic strength $\sigma_{\text{hs}}$ as the two material constants describing the Drucker-Prager strength surface (\ref{eq:DP}). In terms of these, we recall that the uniaxial compressive strength is given by the relation $\sigma_{\text{cs}}=(\sigma_{\text{hs}}/\sigma_{\text{ts}}-2/3)^{-1}\sigma_{\text{hs}}$.




\begin{remark}
  Under states of uniform stress $\stress$ (and hence uniform strain {\bf{E}}) in the body, the phase-field theory (\ref{eq:mombal})-(\ref{eq:ics}) predicts the strength surface 
  \begin{equation}
    \mathcal{F}^{\text{PF}}(\sigma_1,\sigma_2,\sigma_3) = \dfrac{{\rm tr}\,\stress_D^2}{2\mu} + \dfrac{({\rm tr}\,\stress)^2}{9\kappa} - \widehat{c}_e(\mathcal{I}_1, \mathcal{J}_2,0;\ell) - \frac{3\delta^\ell\Gc}{8\ell} = 0.
    \label{eq:strength}
  \end{equation}
  For the case of driving force (\ref{eq:ce}) and coefficient (\ref{eq:delta}), \figref{fig:strength} compares the strength surface \eqref{eq:strength} predicted by the phase-field theory with the actual Drucker-Prager strength surface (\ref{eq:DP}) of a representative material for various values of the regularization length $\ell$.
  
\end{remark}

\section{Spatial and temporal discretizations}
\label{sec:fem}

The governing equations \eqref{eq:mombal} and \eqref{eq:damage_evolution} are second-order hyperbolic and elliptic PDEs, respectively.  Supplemented by their boundary \eqref{eq:bcs} and initial \eqref{eq:ics} conditions, the solution for the displacement field and the phase field can be approximated by finite elements in space and finite differences in time.  


In this work, the displacement field and the phase field are approximated using standard four node quadrilateral elements or eight node brick elements for two and three-dimensional problems, respectively. All simulations are performed using RACCOON~\cite{huRACCOON2022}, a massively parallel finite element code that specializes in phase-field fracture problems. RACCOON is built on the MOOSE framework~\cite{permannMOOSEEnablingMassively2020}, which was developed by Idaho National Laboratory. The momentum balance and phase-field evolution equation are split into subproblems, which are solved using a fixed-point iteration solution scheme, as described by Hu et al.~\cite{huPhasefieldModelFracture2020a}. The phase-field evolution equation is solved using PETSc's variational inequality solver~\cite{bensonFlexibleComplementaritySolvers2006}, with the irreversibility in $\dot{d}$ enforced with a degree-of-freedom-wise primal-dual active-set strategy~\cite{heister2015pdas}.  Importantly, the irreversibility is only enforced once the phase field satisfies $d > 0.95$.  

In addition,  an adaptive mesh refinement strategy is employed to efficiently resolve the pertinent length scales in the vicinity of the regularized fracture surfaces.  
At the start of each time step, the subdomain $\Omega_\mathrm{refine}=\{\mathbf{X}\in\Omega \mid \mathcal{F}(\stress(\mathbf{X}))> -k_1\}$ is refined, where $k_1$ is a small positive value. The expression $\mathcal{F}(\stress(\mathbf{X}))$ serves as an indicator to measure proximity to the strength surface. If it approaches a level where the material strength may soon be violated (i.e., $-k_1$), the corresponding region is refined. This criterion ensures that regions where cracks nucleate and propagate are sufficiently refined.

For the temporal discretization, the HHT-alpha scheme \cite{hughes_2000_hht} is adopted. All simulations are conducted using the parameters $\alpha = -0.25, \beta=0.39$, and $\gamma=0.75$ for which the HHT-alpha method is second-order accurate and unconditionally stable. 

\section{Benchmark problems}
\label{sec:numer}

In the sequel, we make use of the phase-field theory put forth in Section~\ref{sec:form} to work out results for several benchmark problems in dynamic fracture. To aid their analyses, wherever appropriate, we include the corresponding predictions from the cohesive phase-field formulation described in Geelen et al.~\cite{geelen2019} and those obtained using a constitutive prescription for the external driving force $c_e$ different from (\ref{eq:ce}), as described in Kumar et al.~\cite{kumar2020}. 

The models share several common features, for example, they all employ an \text{AT}$_1$ surface regularization ($w(d)=d$) and they all exhibit a threshold for crack nucleation that is largely insensitive to the regularization length. Among their differences, we note that the cohesive model of Geelen et al.~\cite{geelen2019} interprets $d$ as a damage field, and as such enforces monotonicity in $d$, regardless of its magnitude.  By contrast, the phase-field models with strength interpret $d$ as simply an indicator function for the (sharp) crack geometry. As such, the monotonicity in $d$ is only enforced once it exceeds a threshold.  Furthermore, only the cohesive model employs a split of the energy. Results will be shown for this model using both the spectral split \cite{miehe2010thermo}  and the volumetric/deviatoric split \cite{amor2009}. As noted in the Introduction, their most important difference concerns their ability to accurately represent arbitrary strength surfaces for finite regularization lengths $\ell$, as detailed below. The sensitivity of simulated crack nucleation and propagation to these differences is thoroughly examined in this section. 

For clarity, the labels used to refer to the various models in the figures are listed in Table~\ref{tab:models}, alongside some of their features.




\begin{table}[t!]
\small
  \centering
  \begin{tabular}{l|cccccc}
    \toprule
    Model                                & Label   & $g(d)$                                                        & $w(d)$ & Energy Split & Irreversibility            \\
    \midrule
    Cohesive  \cite{geelen2019}          & \coh    & $\dfrac{(1-d)^2}{(1-d)^2 + \tfrac{3\Gc}{8\ell\psi_c} (1+d)d}$ & $d$    & Spectral or      & $\dot{d} >0$               \\[-10pt]
                                         &         &                                                               &        & Vol./Dev. &    \\
    Driving force  $c_e$ \cite{kumar2020}         & \nuc    & $(1-d)^2$                                                     & $d$    & None          & $\dot{d} >0$ when $d>0.95$ \\
    Driving force $c_e$  \cite{kamarei2024} & \latest & $(1-d)^2$                                                     & $d$    & None          & $\dot{d} >0$ when $d>0.95$ \\
    \bottomrule
  \end{tabular}
  \caption{Phase-field models used in the numerical simulations.} \label{tab:models}
\end{table}

\subsection{Single edge notch test: Insensitivity to the regularization length $\ell$}

\begin{figure}[b!]
  \centering
  \begin{subfigure}[b]{0.45\columnwidth}
    \centering
    \includegraphics[height=2.6in]{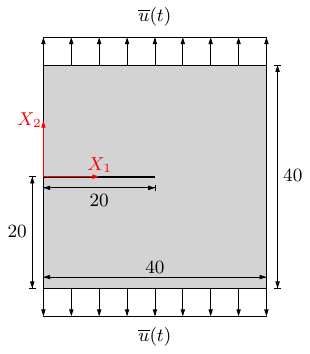}
    \caption{}
  \end{subfigure}
  \hfill
  \begin{subfigure}[b]{0.45\columnwidth}
    \centering
    \includegraphics[width=\columnwidth]{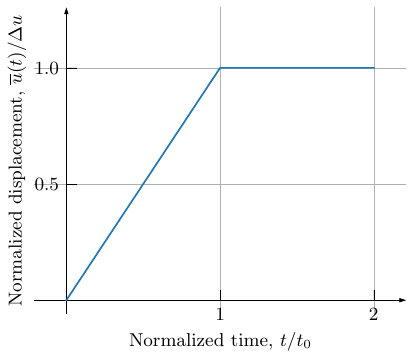}
    \caption{}
  \end{subfigure}
  \caption{(a) Schematic of the single edge notched test (dimenssions in mm). (b) Displacement profile applied at the top and bottom boundaries of the specimen.}
  \label{fig:straight_scheme}
\end{figure}

Consider a square plate with an initially horizontal edge crack as shown in \figref{fig:straight_scheme}(a). The plate occupies the domain $\Omega=[0, 40]~$mm $\times$ $[-20, 20]~$mm. The plate is assumed to be initially at rest, then a vertical displacement is applied at the top and bottom boundaries, as depicted in the figure. The applied displacement first increases linearly and then remains constant, as shown in \figref{fig:straight_scheme}(b), where $t_0=20~\mu$s, $\Delta u = 0.0015~$mm. The material properties and model parameters used in the simulations are given in \tbref{tab:straight_mat}. For all simulations, which are carried out under plane-stress conditions, the problem is discretized in space with an adaptively refined mesh and a time step of 0.1 $\mu$s. 

\begin{table}[t!]
  \centering
  \begin{tabular}{l|cc|c}
    \toprule
    Material properties             & Symbol        & Unit     & Value                                          \\
    \midrule
    Young's modulus                 & $E$           & GPa      & 32                                              \\
    Poisson's ratio                 & $\nu$         & /        & 0.2                                             \\
    Density                         & $\rho$        & g/cm$^3$ & 2.45                                            \\
    Critical energy release rate    & $\Gc$         & N/mm     & 0.003                                           \\
    Uniaxial tensile strength       & $\sigma_{\text{ts}}$ & MPa      & 5                                             \\
    Uniaxial compressive strength   & $\sigma_{\text{cs}}$ & MPa      & 18                                             \\
    \midrule
    Parameters                      &               &          & \latest \\
    \midrule
    Regularization length & $\ell$        & mm       & 0.375,~0.5,~0.625              \\
    Coarse mesh size                & $h_c$         & mm       & 1                  \\
    Fine mesh size                  & $h_f$         & mm       & 0.0625             \\
    \bottomrule
  \end{tabular}
  \caption{Material properties of and model parameters used in the simulations of the single edge notched tests.}
  \label{tab:straight_mat}
\end{table}

\begin{figure}[htbp!]
  \centering
  \includegraphics[width=0.97\columnwidth]{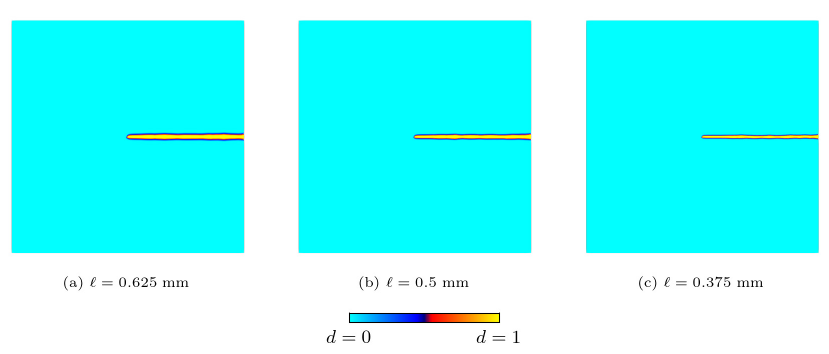}
  \caption{Contour plots of the phase field $d$ resulting from simulations of the single edge notched test with a series of regularization lengths.}
  \label{fig:straight_pf}
\end{figure}

\begin{figure}[htbp!]
  \centering
  \includegraphics[width=.97\columnwidth]{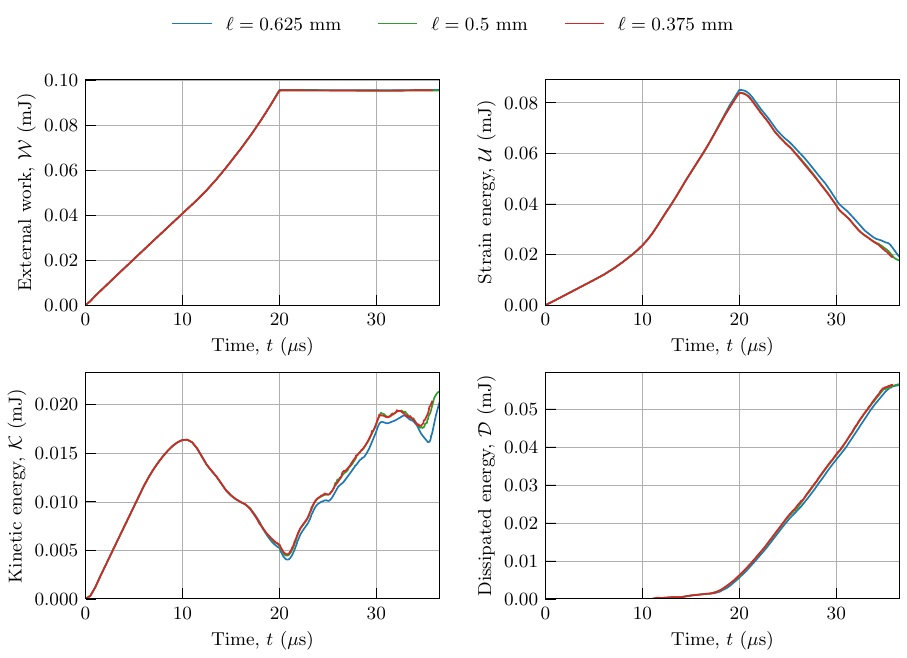}
  \caption{Evolution of the various energies during single-edge notched tests: external work $\mathcal{W}$, strain energy $\mathcal{U}$, kinetic energy $\mathcal{K}$, and dissipated energy $\mathcal{D}$.}
  \label{fig:straight_energy}
\end{figure}

The final fracture patterns for three different values of the regularization length $\ell$ are shown in \figref{fig:straight_pf}, where a single straight crack is observed. 

We now analyze the tests from an energetic perspective, measuring the external work, strain energy, kinetic energy, and dissipated energy throughout the simulation.
In view of the presence of the external driving force $c_e$ in the proposed phase-field model, we evaluate the dissipated energy via
\begin{equation}
  \mathcal{D} = \mathcal{W} - \mathcal{K} - \mathcal{U},
\end{equation}
where $\mathcal{W}$ denotes the work done by the external forces, and the kinetic energy $\mathcal{K}$ and the strain energy $\mathcal{U}$ are defined by
\begin{align}
  \mathcal{K}  = \int_\body \frac 1 2 \rho \dot{\bf{u}}\cdot\dot{\bf{u}}\, {\rm d}{\bf{X}}\quad {\rm and}\quad
  \mathcal{U}  = \int_\body g(d) W (\strain)\,  {\rm d}{\bf{X}}.
\end{align}

A comparison of the external work, the strain energy, the kinetic energy, and the dissipated energy in single edge notched tests across three different values of regularization length is shown in \figref{fig:straight_energy}. Results for regularization lengths smaller than $0.5$~mm are almost indistinguishable.


\subsection{The Kalthoff--Winkler experiment: The essential role of the strength surface of the material}

\begin{figure}[t!]
  \centering
  \includegraphics[width=.45\columnwidth]{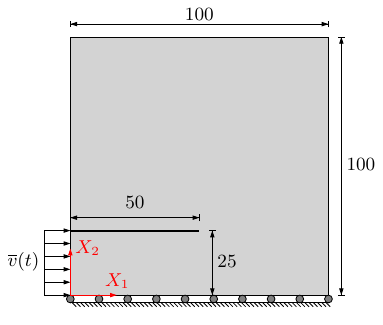}
  \caption{Domain and boundary conditions for the Kalthoff-Winkler problem (dimensions in mm).}
  \label{fig:kal_scheme}
\end{figure}

We now examine a second benchmark problem, one of the classical experiments by Kalthoff and Winkler \cite{kalthoff1988}, wherein a projectile impacts a plate with two edge notches. We focus on the version of the experiments with low-impact velocities. In the results that follow, symmetry is assumed and only the upper half of the specimen is explicitly modeled. As shown by \figref{fig:kal_scheme}, the computational domain is taken to be $\Omega=[0, 100]~$mm $\times$ $[0, 100]~$mm. The plate is assumed to be initially at rest. To approximate the loading from the impact, we apply a horizontal velocity to the notched area.  The velocity is applied as a boundary condition that is ramped up linearly to a magnitude of $v_0$ and then held constant:
\begin{equation}
  \overline{v}_1(X_1=0, X_2\in[0, 25], t) =
  \begin{cases}
    \dfrac{v_0}{t_0} t & \text{if } t < t_0   \\
    v_0                & \text{if } t \ge t_0
  \end{cases}
\end{equation}
with $v_0=16.5$~m/s and $t_0=1~\mu s$.  In this work, this is enforced with a
Dirichlet boundary condition on the displacement field given by
\begin{equation}
  \overline{u}_1(X_1=0, X_2\in[0, 25], t) =
  \begin{cases}
    \dfrac{v_0}{2 t_0} t^2      & \text{if } t < t_0   \\
    v_0 t - \dfrac{1}{2}v_0 t_0 & \text{if } t \ge t_0
  \end{cases}.
\end{equation}
The material properties and model parameters used in the simulations are given in \tbref{tab:kal_mat}. All simulations are carried out under plane-stress conditions using a time step of 0.5~$\mu$s.

\begin{table}[htbp!]
  \centering
  \begin{tabular}{l|cc|ccc}
    \toprule
    Material properties           & Symbol        & Unit     & \multicolumn{3}{c}{Value}                                         \\
    \midrule
    Young's modulus               & $E$           & GPa      & \multicolumn{3}{c}{190}                                           \\
    Poisson's ratio               & $\nu$         & /        & \multicolumn{3}{c}{0.3}                                           \\
    Density                       & $\rho$        & g/cm$^3$ & \multicolumn{3}{c}{8}                                             \\
    Critical energy release rate  & $\Gc$         & N/mm     & \multicolumn{3}{c}{22.2}                                          \\
    Nucleation energy             & $\psi_c$      & MPa      & \multicolumn{3}{c}{7.9}                                           \\
    Uniaxial tensile strength     & $\sigma_{\mathrm{ts}}$ & MPa      & \multicolumn{3}{c}{1733}                                          \\
    Uniaxial compressive strength & $\sigma_{\mathrm{cs}}$ & MPa      & \multicolumn{3}{c}{5500}                                          \\
    \midrule
    Parameters                    &               &          & \coh          & \nuc & \latest \\
    \midrule
    Regularization length         & $\ell$        & mm       & 0.75                      & 0.75             & 0.75               \\
    Coefficient                   & $\delta^\ell$ & /        & /                         & 0                & 0                  \\
    Coarse mesh size              & $h_c$         & mm       & 5                         & 5                & 5                  \\
    Fine mesh size                & $h_f$         & mm       & 0.15625                   & 0.15625          & 0.15625            \\
    \bottomrule
  \end{tabular}
  \caption{Material properties of steel and model parameters used in the simulations of the Kalthoff-Winkler experiments.}
  \label{tab:kal_mat}
\end{table}

\begin{figure}[htbp!]
  \centering
  \includegraphics[width=0.97\columnwidth]{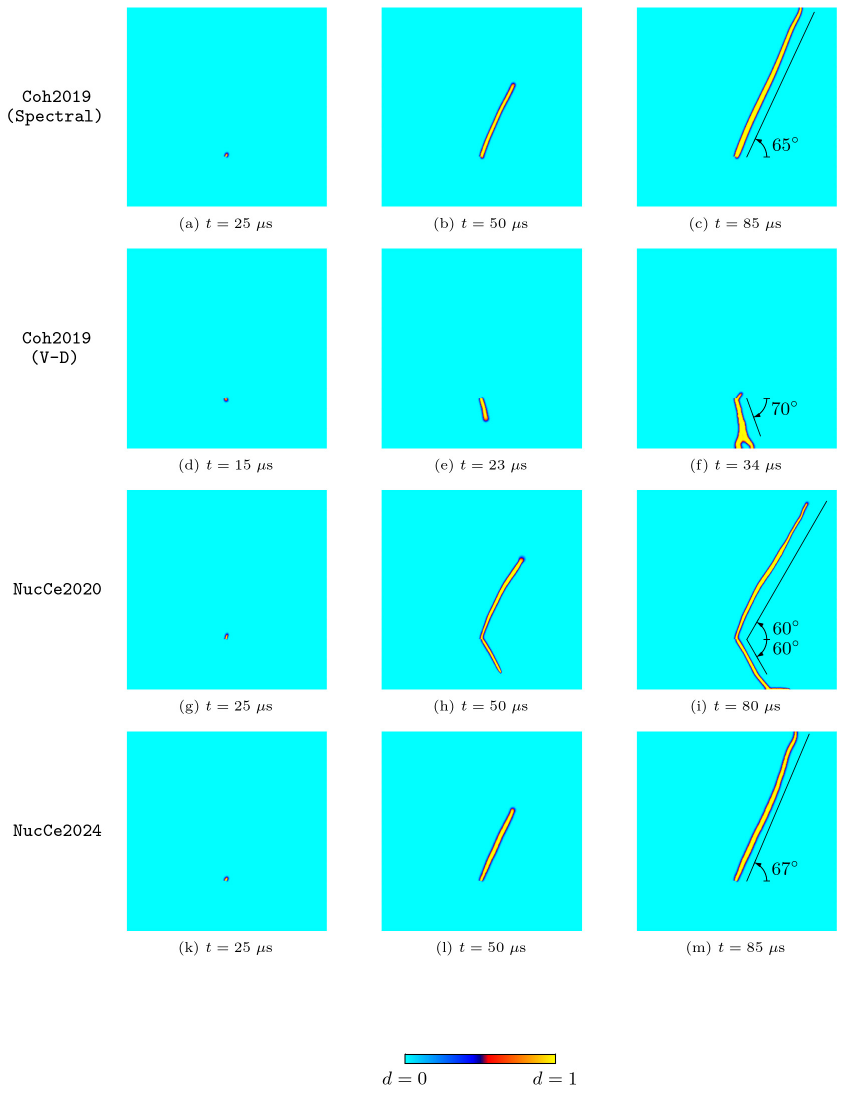}
  \caption{Contour plots of the phase field $d$ in the Kalthoff-Winkler experiments from \coh~\texttt{(Spectral)}, \coh~\texttt{(VD)}, \nuc, and \latest~at selected time steps.}
  \label{fig:kal_pf}
\end{figure}

\begin{figure}[htbp!]
  \centering
  \includegraphics[height=.3735\columnwidth]{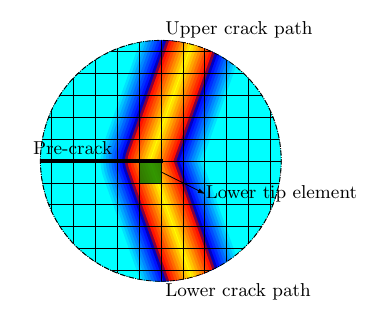}
  \caption{Schematic of crack paths near the tip of the initial notch in the Kalthoff-Winkler simulations. The lower tip element (marked blue) and upper tip element (marked green) are the first cracked element of the lower crack path and the upper crack path in the phase-field simulations shown in \figref{fig:kal_pf}.
  }
  \label{fig:kal_tip}
\end{figure}

\begin{figure}[htbp!]
  \centering
  \includegraphics[width=.97\columnwidth]{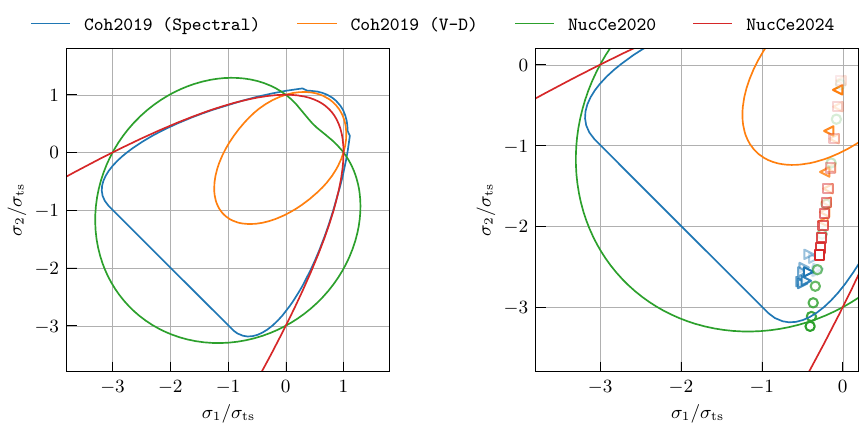}
  \caption{Two-dimensional contours of the  strength surfaces (for $\sigma_3=0$)  corresponding to the various phase-field models used to simulate the Kalthoff-Winkler problem (left), using the material properties and parameters in Table~\ref{tab:kal_mat}. The scatter points in the zoomed view (right) of the third quadrant indicate the evolution of the stress state at the lower crack tip element $(X_1,X_2)=(50, 24.9)$ mm (see \figref{fig:kal_tip}) over the time interval $t\in[10, 30]~\mu$s, up to the point of crack nucleation.  The scatter points are colored to match the various models, as listed in the legend across the top of the plots.
  }
  \label{fig:kal_strength}
\end{figure}


\begin{figure}[htbp!]
  \centering
  \includegraphics[width=.97\columnwidth]{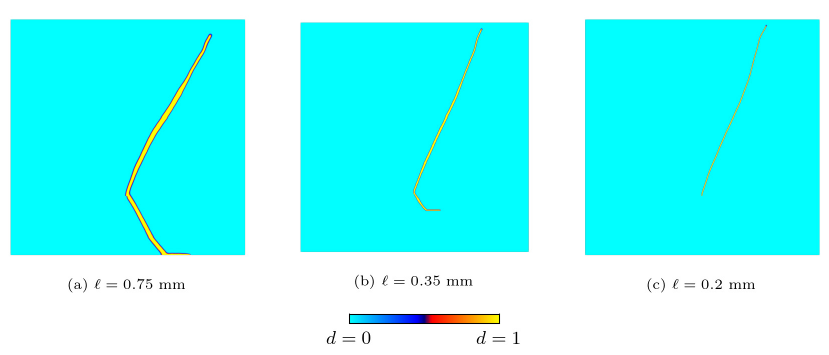}
  \caption{Contour plots of the phase field $d$ resulting from \nuc\, at $t=80~\mu$s with a series of regularization lengths.}
  \label{fig:kal_nuc20_pf}
\end{figure}

Contours of the phase field  generated by the various models are shown in \figref{fig:kal_pf}, at select points in time. The results from the cohesive model with a volumetric-deviatoric split (referred to \coh~\texttt{(V-D)} are shown in \figref{fig:kal_pf}(d-f). The results from the original cohesive model with the spectral split, denoted as \coh~\texttt{(Spectral)}, are shown in \figref{fig:kal_pf}(a-c). This comparison elucidates the clear influence of the  strength surface on dynamic fracture. 

In the experimental observations by Kalthoff and Winkler \cite{kalthoff1988}, the angle between the final crack path and the initial notch was approximately $70^\circ$. Compared to the experimental results, the cohesive model with spectral split and the current nucleation model yield reasonable crack paths with angles of approximately $65^\circ$ and $67^\circ$, respectively. However, the results obtained using the cohesive model with a volumetric-deviatoric split and the original nucleation model indicate spurious crack patterns, with cracks growing at negative angles from the initial notch.

Some insight into these results can be obtained by examining the principal stresses in an element near the tip of the initial notch. As shown in \figref{fig:kal_tip}, both upper and lower crack paths can form in the results.  The figure identifies a ``lower tip element" that is located just under the tip of the notch.  \figref{fig:kal_strength} shows the effective strength surfaces in principal stress space for the various models. On the right, the stress trajectories in the lower tip element $(X_1,X_2)= (50, 24.9)$  are indicated by scatter points during the time just prior to crack nucleation, i.e.,\  $t\in [10, 30]~\mu$s.  The principal stresses in this lower tip element happen to be in the third quadrant of principal stress space, as the zoom on the right of \figref{fig:kal_strength} illustrates.  In this region, the cohesive model with the volumetric-deviatoric split has the smallest strength surface, and it is quickly violated.  This appears to be the explanation for the crack pattern shown in \figref{fig:kal_pf}(e) and \figref{fig:kal_pf}(f).

It bears emphasis that the driving force $c_e$ plays two important roles.  Under uniform states of stress, it 
enforces the condition that crack nucleation cannot occur until the strength surface is exceeded.  But for large pre-existing cracks, is also prevents crack nucleation under compression.  This latter role is clearly illustrated in this problem by comparing the results generated by the \nuc \, and \latest \,  models.  The two models are identical, with the exception of how accurately they represent the strength surface for the same finite value of $\ell$.  What can be observed from the results shown in \figref{fig:kal_pf}(g-m) and \figref{fig:kal_tip} is that material points exceed the strength surface in this lower tip element for the \nuc \, model, but not the \latest \, model.  As such, the latter prevents the nucleation and growth of the spurious branch in compression, and yields only a crack that propagates up and to the right, consistent with the experimental observations. As noted in \cite{kumar2020}, the effective strength surface for the \nuc\, model asymptotes to the Drucker-Prager strength surface as the regularization length decreases. As shown in \figref{fig:kal_nuc20_pf}, the spurious branch vanishes when the regularization length is reduced to 0.2~mm.

The \coh~\texttt{(Spectral)} is somewhat of an outlier here, in the sense that its effective strength surface is quite close to that of the \nuc \, model, but material points do not quite exceed it.  This is due to the remaining differences between the models, such as the degradation functions and how irreversibility is treated.  What is observed with this model is that some damage does indeed develop below the existing edge crack, but it does not fully develop into a lower branch.  As a result, the upper branch is allowed to develop and generally evolve with the correct orientation.


\subsection{Fragmentation of a hollow sphere under internal pressure: Robustness of the numerical scheme}

\begin{figure}[t!]
  \centering
  \begin{subfigure}[b]{0.45\columnwidth}
    \centering
    \includegraphics[height=0.62\columnwidth]{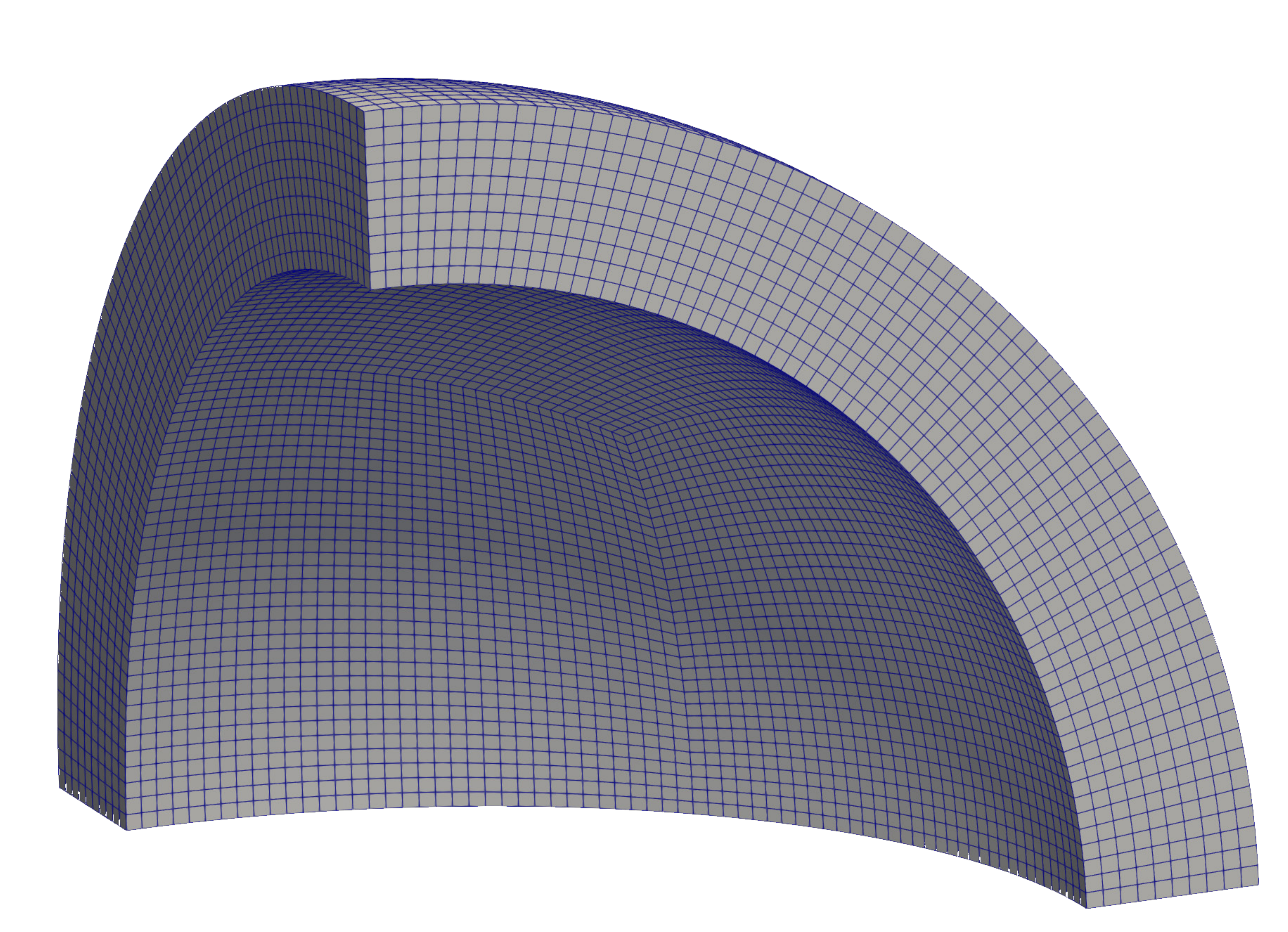}
    \caption{}
  \end{subfigure}
  \hfill
  \begin{subfigure}[b]{0.45\columnwidth}
    \centering
    \includegraphics[height=0.62\columnwidth]{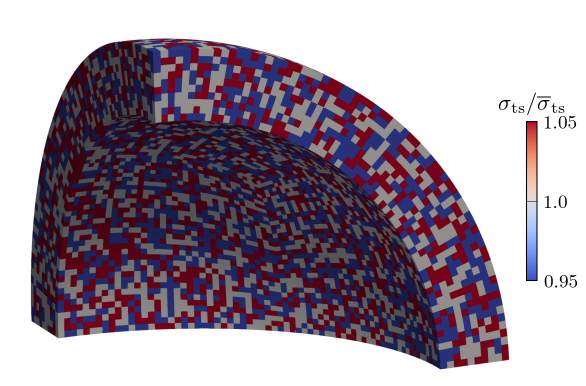}
    \caption{}
  \end{subfigure}
  \caption{(a) The initial mesh used to simulate the fragmentation of a hollow sphere, and (b) a spatially random tensile strength field.}
  \label{fig:frag_scheme}
\end{figure}

We now investigate a problem that involves dynamic fragmentation, illustrating crack nucleation and growth in three dimensions. Consider a hollow sphere subjected to an impulsive pressure on its inner boundary.\footnote{A related problem, the fragmentation of a cylinder, has been studied by both Geelen et al.~\cite{geelen2019} and \cite{WuDynamic2020} using cohesive phase field models.}  As shown in \figref{fig:frag_scheme}(a), only 1/8th of the hollow sphere geometry is explicitly modeled, and symmetry conditions are imposed on the flat cut surfaces. The sphere has inner and outer radii of 80 and 100~mm, respectively. The pressure enforced on the inner boundary takes the form $p_0(t) = p_0 \exp{\left(-t/t_0\right)}$, with an initial pressure of $p_0=400$~MPa and a decay time of $t_0=30~\mu$s.  A time step of 0.5~$\mu$s is used for the simulations of this problem.

The material properties used are listed in Table~\ref{tab:frag_mat}. The
domain is partitioned using a structured mesh of 8-node brick elements, as shown in \figref{fig:frag_scheme}(a). As the domain and applied loading gives rise to stress fields that are spatially uniform at any given radius, the model must be perturbed in some manner to help seed the localization of fractures. Accordingly, we introduce a spatially random field for the tensile strength and the hydrostatic strength, as shown in \figref{fig:frag_scheme}(b). The tensile and hydrostatic strengths vary from the baseline material properties by $\sigma_{\text{ts}}/\overline{\sigma}_{\text{ts}} = \{0.95, 1, 1.05 \}$ over patches of elements that are approximately five times larger than the regularization length $\ell$.  Finally, to avoid large distortions and interpenetration of fully damaged elements, we decrease the magnitude of the applied pressure on the inner surface with the degradation function as
\begin{equation}
  p({\bf{X}}, t)=g(d({\bf{X}}, t))p_0\exp{\left(-t/t_0\right)}.
\end{equation}

\begin{table}[t!]
  \centering
  \begin{tabular}{l|cc|c}
    \toprule
    Material properties                    & Symbol                   & Unit     & Value              \\
    \midrule
    Young's modulus                        & $E$                      & GPa      & 210                \\
    Poisson's ratio                        & $\nu$                    & /        & 0.3                \\
    Density                                & $\rho$                   & g/cm$^3$ & 7.85               \\
    Critical energy release rate           & $\Gc$                    & N/mm     & 20                 \\
    Baseline uniaxial tensile strength     & $\overline{\sigma}_{\text{ts}}$ & MPa      & 1000               \\
    Baseline uniaxial hydrostatic strength & $\overline{\sigma}_{\text{hs}}$ & MPa      & 8000               \\
    Rayleigh wave speed                    & $c_R$                    & m/s      & 2971               \\
    \midrule
    Parameters                             &                          &          & \latest \\
    \midrule
    Regularization length                  & $\ell$                   & mm       & 1                  \\
    Coarse mesh size                       & $h_c$                    & mm       & 2                  \\
    Fine mesh size                         & $h_f$                    & mm       & 0.25               \\
    \bottomrule
  \end{tabular}
  \caption{Material properties and model parameters used in the fragmentation simulations.}
  \label{tab:frag_mat}
\end{table}

\begin{figure}[t!]
  \centering
  \includegraphics[width=.97\columnwidth]{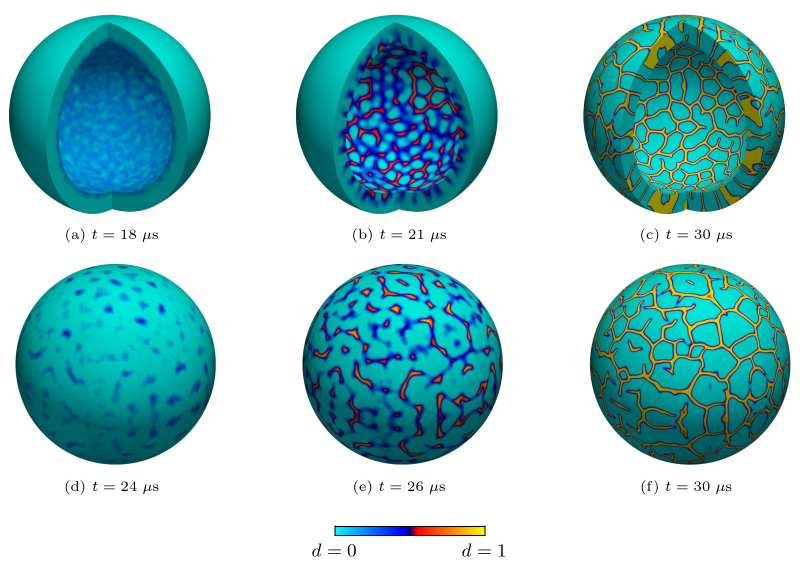}
  \caption{Contour plots of the phase field $d$ for a pressurized, hollow sphere using the \latest \, model, viewed from the interior (a)-(c) and exterior (d)-(f) of the sphere at selected time steps.}
  \label{fig:frag_pf}
\end{figure}

\begin{figure}[t!]
  \centering
  \includegraphics[width=.41\columnwidth]{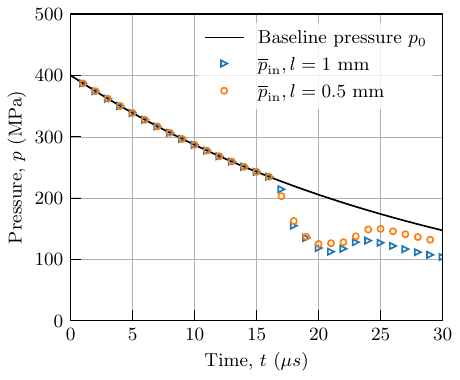}
  \caption{Comparison of the average degraded pressure and the baseline pressure on the inner spherical surface, as a function of time, for simulations of the pressurized hollow sphere problem with two different regularization lengths.}
  \label{fig:frag_pressure}
\end{figure}

\figref{fig:frag_pf} shows the resulting fracture evolution at various times, viewed from the interior and exterior of the hollow sphere. Cracks start nucleating at about $t=18~\mu$s with a spatially varying small phase field on the inner spherical boundary, then localize to cracks within roughly 3~$\mu$s. Some cracks reach the outer boundary at approximately $t=24~\mu$s, and the fragmentation process completes at approximately $t=30~\mu$s.

To quantify the degraded pressure on the inner boundary, we define the surface average degraded pressure as
\begin{equation}
  \overline{p}_{\text{in}}({\bf{X}}, t) = \frac{1}{\text{Area}({S_\text{in}})} \int_{S_{\text{in}}} p({\bf{X}}, t) \; \mathrm{d}{\bf{X}}.
\end{equation}
\figref{fig:frag_pressure} compares the magnitude of the surface average degraded pressure on the inner spherical surface to the baseline pressure as a function of time, for two different regularization lengths.  As expected, prior to any crack nucleation, all pressures are identical, and the difference between the pressures diminishes as the regularization length if decreased.  The decrease in distance between the pressure curves after $t=20~\mu$s results from the fact that the irreversibility in the phase field is only enforced for $d > 0.95$.

\section{Comparison with experiments}\label{sec:experiments}

\subsection{Dynamic Brazilian fracture experiments on Basalt}

\begin{figure}[h!]
  \centering
  \begin{subfigure}[b]{0.45\columnwidth}
    \centering
    \includegraphics[]{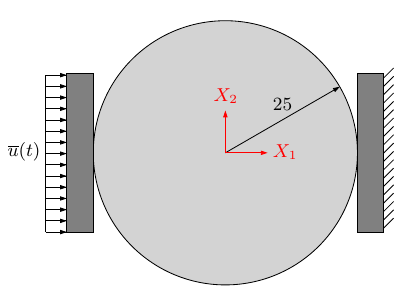}
    \caption{}
  \end{subfigure}
  \hfill
  \begin{subfigure}[b]{0.45\columnwidth}
    \centering
    \includegraphics[height=0.68\columnwidth]{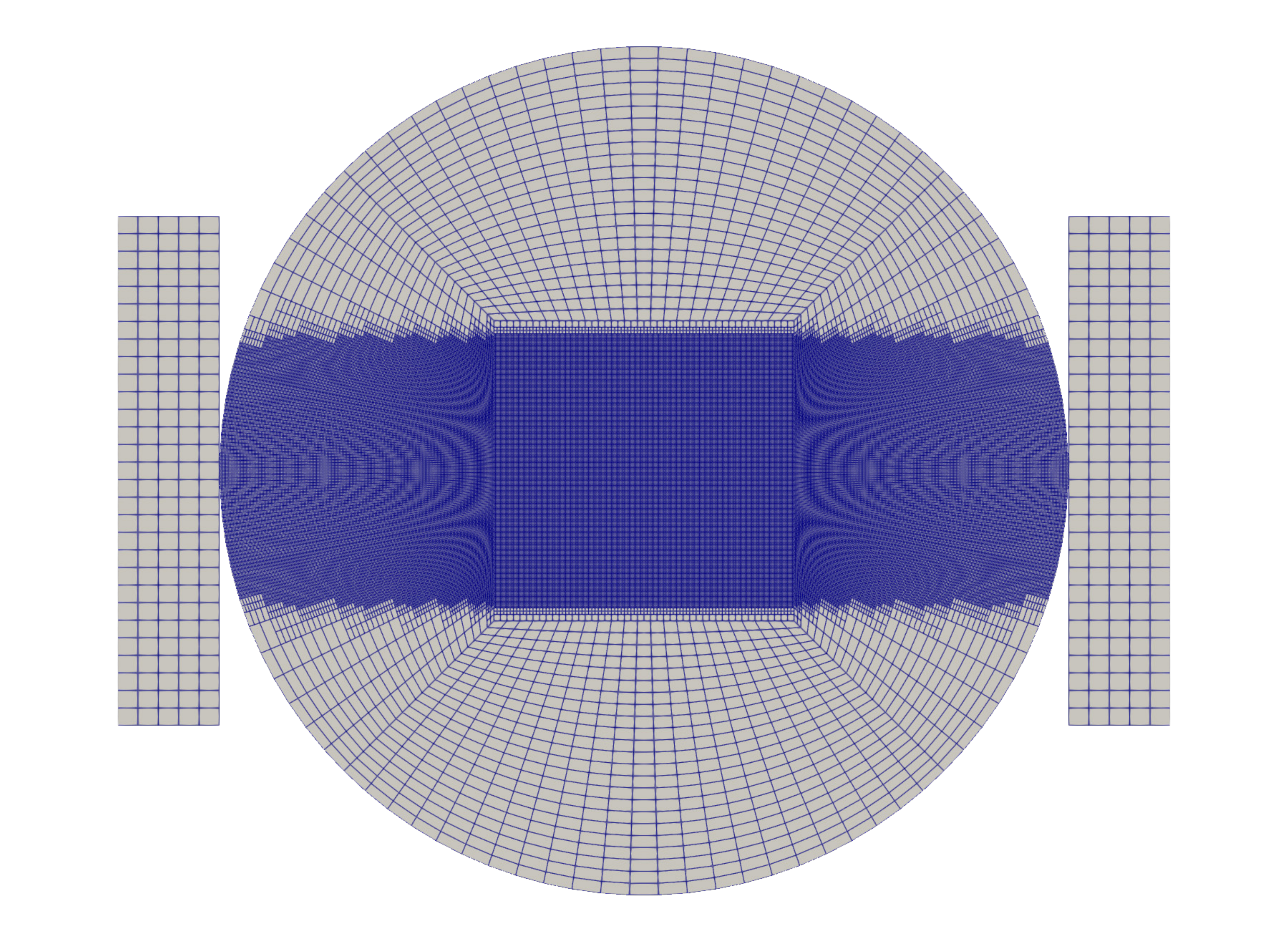}
    \caption{}
  \end{subfigure}
  \caption{(a) Schematic of the dynamic Brazilian fracture experiments (dimensions in mm), and (b) the finite element mesh.}
  \label{fig:brz_scheme_mesh}
\end{figure}


We now turn to simulating the dynamic Brazilian experiments on Basalt reported by Yin et al.~\cite{yin2022dynamic_brz}. In these experiments, a split Hopkinson pressure bar setup was used to dynamically load a disk of Basalt in compression.  In the simulations which follow, the specimen is taken to be initially at rest, and plane-stress conditions are assumed.  The impact of the long bar is approximated by imposing contact conditions between the disk and a rectangular, rigid domain with a prescribed displacement; see \figref{fig:brz_scheme_mesh}.  Contact conditions are also imposed between the specimen and a rigid block on its right side, whose displacement is held fixed.  Both conditions effectively approximate impact and bars that are much stiffer than the specimen.   As shown in \figref{fig:brz_scheme_mesh}(b), the quadrilateral mesh used in the simulations is highly refined around the center of the specimen, where crack nucleation and growth was observed in the experiments. For this problem, a time step of 0.5~$\mu$s is used.

The horizontal component of the displacement field for the rigid left bar is prescribed as
\begin{equation}
  \overline{u}(t) = \frac{u_0}{2}\left(-\cos{\left(\frac{\pi t}{t_0} \right)} + 1\right),
\end{equation}
where $u_0=1$~mm and $t_0=100~\mu$s. The magnitude of this prescribed displacement  was determined by matching the stress at fracture observed in the experiments to an elastodynamic simulation (without any modeling of fracture).  

\begin{table}[t!]
  \centering
  \begin{tabular}{l|cc|ccc}
    \toprule
    Material properties             & Symbol        & Unit     & \multicolumn{3}{c}{Value}                                          \\
    \midrule
    Young's modulus                 & $E$           & GPa      & \multicolumn{3}{c}{20.11}                                          \\
    Poisson's ratio                 & $\nu$         & /        & \multicolumn{3}{c}{0.2}                                            \\
    Density                         & $\rho$        & g/cm$^3$ & \multicolumn{3}{c}{2.74}                                           \\
    Critical energy release rate    & $\Gc$         & N/mm     & \multicolumn{3}{c}{0.1}                                            \\
    Nucleation energy               & $\psi_c$      & MPa      & \multicolumn{3}{c}{0.0032}                                         \\
    Uniaxial tensile strength       & $\sigma_{\text{ts}}$ & MPa      & \multicolumn{3}{c}{11.31}                                          \\
    Uniaxial compressive strength   & $\sigma_{\text{cs}}$ & MPa      & \multicolumn{3}{c}{339.3}                                          \\
    \midrule
    Parameters                      &               &          & \coh           & \nuc & \latest \\
    \midrule
    Regularization length & $\ell$        & mm       & 1.25                       & 1.25             & 1.25               \\
    Coarse mesh size                & $h_c$         & mm       & 0.75                       & 0.75             & 0.75               \\
    Fine mesh size                  & $h_f$         & mm       & 0.1                        & 0.1              & 0.1                \\
    \bottomrule
  \end{tabular}
  \caption{Material properties of Basalt and model parameters used in the simulations of the dynamic Brazilian fracture experiments.}
  \label{tab:brz_mat}
\end{table}

\begin{figure}[htbp!]
  \centering
  \includegraphics[width=.9\columnwidth]{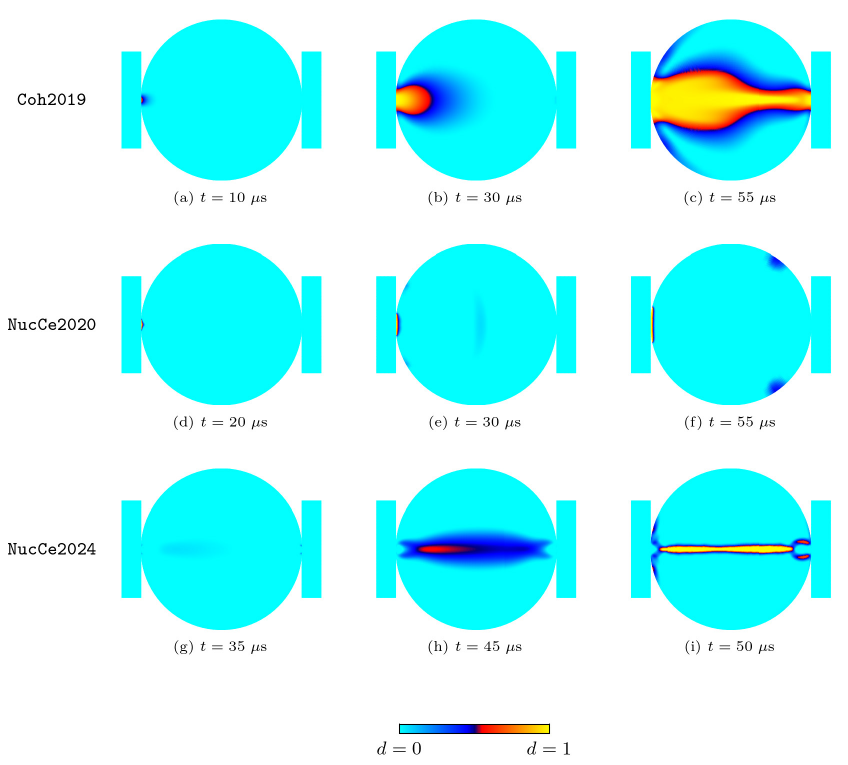}
  \caption{Contour plots of the phase field $d$ from simulations of dynamic Brazilian fracture experiments using a cohesive model with a spectral split (top row), the \nuc\, model (center row), and the \latest\, model (bottom row), at selected time steps.}
  \label{fig:brz_pf}
\end{figure}

\begin{figure}[htbp!]
  \centering
  \begin{subfigure}[b]{\columnwidth}
    \centering
    \includegraphics[width=0.99\columnwidth]{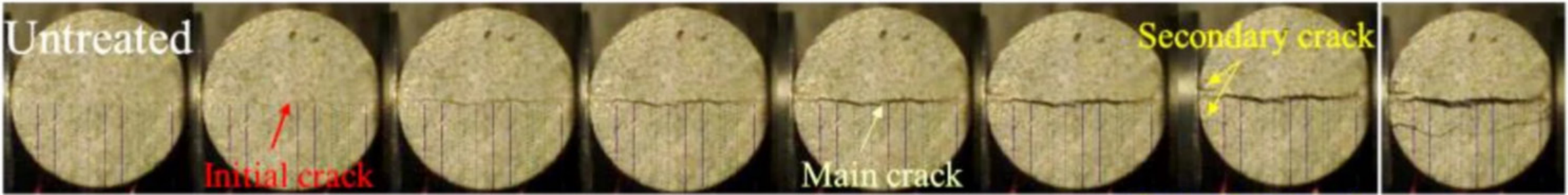}
    \caption{}
  \end{subfigure}
  \begin{subfigure}[b]{\columnwidth}
    \centering
    \includegraphics[width=\columnwidth]{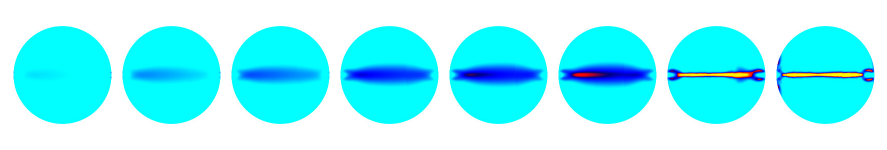}
    \caption{}
  \end{subfigure}
  \caption{Comparison of (a) experimental results from \cite{yin2022dynamic_brz} to (b) phase field contours from the \latest \, model over the time interval $t\in[35, 50]~\mu$s.  The results are shown in sequential order, from left to right.}
  \label{fig:brz_pf_vs_exp}
\end{figure}

\begin{figure}[htbp!]
  \centering
  \includegraphics[width=.99\columnwidth]{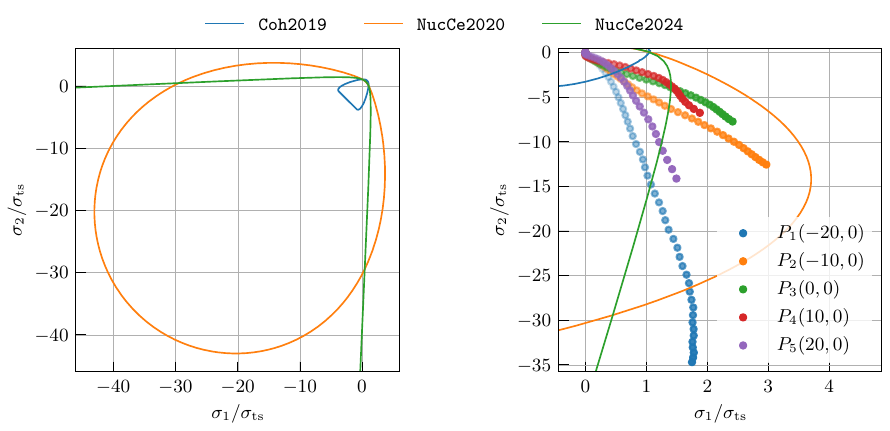}
  \caption{Two-dimensional contours of effective strength surfaces (for $\sigma_3=0$) for Basalt (left) corresponding to various phase field models, using the material properties and parameters listed in \tbref{tab:brz_mat}. The scatter points in the zoom of the fourth quadrant (right) indicate the evolution in the stress state at $P_1=(-20, 0)$, $P_2=(-10, 0)$, $P_3=(0, 0)$, $P_4=(10, 0)$, and $P_5=(20, 0)$ during an elastodynamic simulation (without any phase-field coupling).
  }
  \label{fig:brz_strength}
\end{figure}


Table~\ref{tab:brz_mat} lists the material properties and model parameters used in the simulations of this experiment.  Of note is the relative magnitude of the uniaxial compressive strength to the uniaxial tensile strength. We once again find it illustrative to compare results obtained using the various phase-field models.  \figref{fig:brz_pf}(a-c) reveal that the fracture pattern predicted by the cohesive model with a spectral split is very diffuse, even with a relatively small regularization length of $\ell = 1.25$~mm. In addition, the crack initiates directly from the contact region between the left plate and the specimen. Similarly, as \figref{fig:brz_pf}(d-f) indicate, the crack initiation modeled by the \nuc\, model occurs at the contact surface, with the damage not propagating further into the specimen. By contrast, the results obtained using the \latest \,  model indicate a crack that initiates within the specimen near $(X_1,X_2)=(-18, 0)$, and then propagates horizontally to the right, as illustrated in \figref{fig:brz_pf}(g-i). Furthermore, the crack evolution predicted by the \latest \, model aligns well with experimental observations from \cite{yin2022dynamic_brz}, as shown in \figref{fig:brz_pf_vs_exp}.


\figref{fig:brz_strength} shows the significant differences between the  strength surfaces corresponding to the three models examined for this problem. Recall that all three surfaces are pinned to the uniaxial tensile strength as represented by the points $(\sigma_{\text{ts}}, 0)$ and $(0,\sigma_{\text{ts}})$ in principal stress space.
The strength surfaces of the \nuc\, and \latest\, models also pass through $(-\sigma_{\text{cs}}, 0)$ and $(0, -\sigma_{\text{cs}})$, corresponding to failure under uniaxial compression.

In the cohesive model, by contrast, the uniaxial compressive strength cannot be independently specified.  Instead, it is a function of Poisson's ratio and the uniaxial tensile strength, precisely:
\begin{equation}
  \sigma_{\text{cs}} = \sqrt{\frac{(1+2\nu)(1-\nu)}{2\nu^2}}\sigma_{\text{ts}} \approx 3.74 \sigma_{\text{ts}}.
\end{equation}
For Basalt, the uniaxial compressive strength is typically 15-30 times greater than the uniaxial tensile strength. Therefore, crack nucleation predicted by the cohesive model occurs earlier than in either of the phase-field models with strength. We note that the strength surface of the proposed phase-field model is the only one open in the third quadrant, where $\sigma_1<0$ and $\sigma_2 <0$. Therefore, only the proposed phase-field model provides a favorable comparison to the experiments. It is the only model where the crack does not initiate directly in the contact region where the stress is mostly compressive. Instead, crack nucleation begins in the middle left of the center of the specimen.

To investigate the location of crack nucleation and the stress states in the specimen, we now report results for this problem from an elastodynamics simulation without any phase-field coupling. The stress states are monitored at five points along the specimen's centerline: $P_1=(-20, 0)$, $P_2=(-10, 0)$, $P_3=(0, 0)$, $P_4=(10, 0)$, and $P_5=(20, 0)$. Stress trajectories over time are plotted on the right in \figref{fig:brz_strength}, and illustrate when the various strength surfaces are violated. For the \nuc \, model, the strength surface is only violated at $P_1$.  This aligns with crack localization at the contact point with this model.   By contrast, the strength surfaces of the other two models are violated by the stress trajectories at all five points, indicating that the crack initiates and propagates horizontally through the entire specimen. However, with the cohesive model, the strength surface is violated across a much larger region, which likely explains the much more diffuse nature of the phase-field in the fully-coupled simulation; see \figref{fig:brz_pf}(a-c).

\subsection{Impact experiments on soda-lime glass}
\begin{figure}[htbp!]
  \centering
  \begin{subfigure}[b]{0.49\columnwidth}
    \centering
    \includegraphics[]{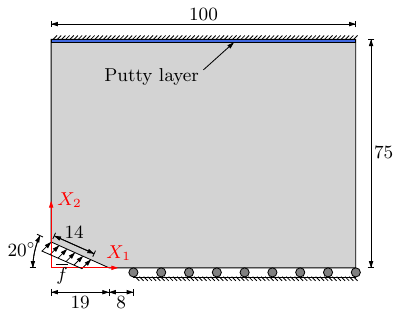}
    \caption{}
  \end{subfigure}
  \hfill
  \begin{subfigure}[b]{0.49\columnwidth}
    \centering
    \includegraphics[]{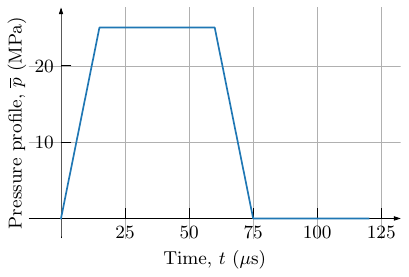}
    \caption{}
  \end{subfigure}
  \caption{(a) Schematic of the soda-lime glass impact experiment (dimensions in mm). (b) Pressure profile applied on the V-notch surfaces.}
  \label{fig:soda_scheme}
\end{figure}

Finally, we investigate recent experimental observations for the dynamic fracture of soda-lime glass reported by Sundaram and Tippur~\cite{sundaram_soda_exp_2018}. In these experiments, soda-lime glass specimens with V-notches and small initial edge cracks were subjected to an impact on the faces of the notch using a Hopkinson bar apparatus.  The entire impact event was reported to occur over a time of approximately 125~$\mu$s. The geometry of the specimen and the loading were sufficient to give rise to straight crack propagation, followed by crack branching.  A comparison of results for simulations of this problem using both phase-field models for fracture and various peridynamics models was reported by Mehrmashhadi et al.~\cite{mehrmashhadi_soda_peridynamic_2020}.  

The geometry and boundary conditions for the model used in this work are depicted in \figref{fig:soda_scheme}(a). Symmetry conditions are used to consider only the top portion of the specimen.   
Previous studies of this problem have made it clear that the results are fairly sensitive to the applied load that is used at the V-notch to model the impact event. Here, we begin by considering the approach described by Mehrmashhadi et al.~\cite{mehrmashhadi2019pmma,mehrmashhadi_soda_peridynamic_2020} in which the results from a 
three-dimensional LS-DYNA model of the impact bar were used to extract a reasonable pressure profile to apply to the faces of the V-notch in their fracture simulations. According to \cite{mehrmashhadi2019pmma}, since the LS-DYNA model did not account for crack propagation, the pressure signal it provided is only valid up to the time of crack initiation.  However, when inertial effects are properly accounted for, information travels at finite speeds.  It takes time for the crack propagation to influence the response at the face of the V-notch.  As such, the pressure pulse should be valid over a slightly longer period of time.  Indeed, we calculated the time required for a wave to propagate from the crack tip to the V-notch, and adjusted the duration of the pressure pulse accordingly.  The pressure profile used in this work is illustrated in \figref{fig:soda_scheme}(b). 

Two types of loads are considered: the first is a pressure load that is strictly normal to the V-notch, and the second also considers a tangential load, which is viewed as accounting for a potential frictional contribution. Assuming a friction coefficient of 0.35 between the glass and the impact bar, the applied traction on the faces of the V-notch then takes one of the two following forms:
\begin{equation}
  \overline{\bs{f}} =
  \begin{cases}
    \overline{p} {\bf{N}}                             & \text{without tangential load} \vspace{0.2cm}\\
    \overline{p} {\bf{N}} + 0.35\overline{p} {\bf{N}}_t & \text{with tangential load}
  \end{cases},
\end{equation}
where ${\bf{N}}$ and ${\bf{N}}_t$ represent the normal and tangential unit vector of the V-notch surface.

Table~\ref{tab:soda_mat} lists the material properties and model parameters used for the soda-lime impact simulations presented in this work. For this problem, plane-stress conditions are assumed, and a time step of 0.5~$\mu$s is used.

\begin{table}[t!]
  \centering
  \begin{tabular}{l|cc|c}
    \toprule
    Material properties of soda-lime glass & Symbol        & Unit     & Value              \\
    \midrule
    Young's modulus                        & $E$           & GPa      & 72                 \\
    Poisson's ratio                        & $\nu$         & /        & 0.25               \\
    Density                                & $\rho$        & g/cm$^3$ & 2.44               \\
    Critical energy release rate           & $\Gc$         & N/mm     & 9                  \\
    Uniaxial tensile strength              & $\sigma_{\mathrm{ts}}$ & MPa      & 30                 \\
    Uniaxial compressive strength          & $\sigma_{\mathrm{cs}}$ & MPa      & 330                \\
    Rayleigh wave speed                    & $c_R$         & m/s      & 3170               \\
    \midrule
    Material properties of putty           &               &          &                    \\
    \midrule
    Young's modulus                        & $E_p$         & MPa      & 1.7                \\
    Poisson's ratio                        & $\nu_p$       & /        & 0.4                \\
    Density                                & $\rho_p$      & g/cm$^3$ & 1                  \\
    \midrule
    Parameters                             &               &          & \latest \\
    \midrule
    Regularization length                  & $\ell$        & mm       & 0.25               \\
    Coarse mesh size                       & $h_c$         & mm       & 1                  \\
    Fine mesh size                         & $h_f$         & mm       & 0.0625             \\
    \bottomrule
  \end{tabular}
  \caption{Material properties and model parameters used in the simulations of the soda-lime impact experiments.}
  \label{tab:soda_mat}
\end{table}

\begin{figure}[t!]
  \centering
  \includegraphics[width=.99\columnwidth]{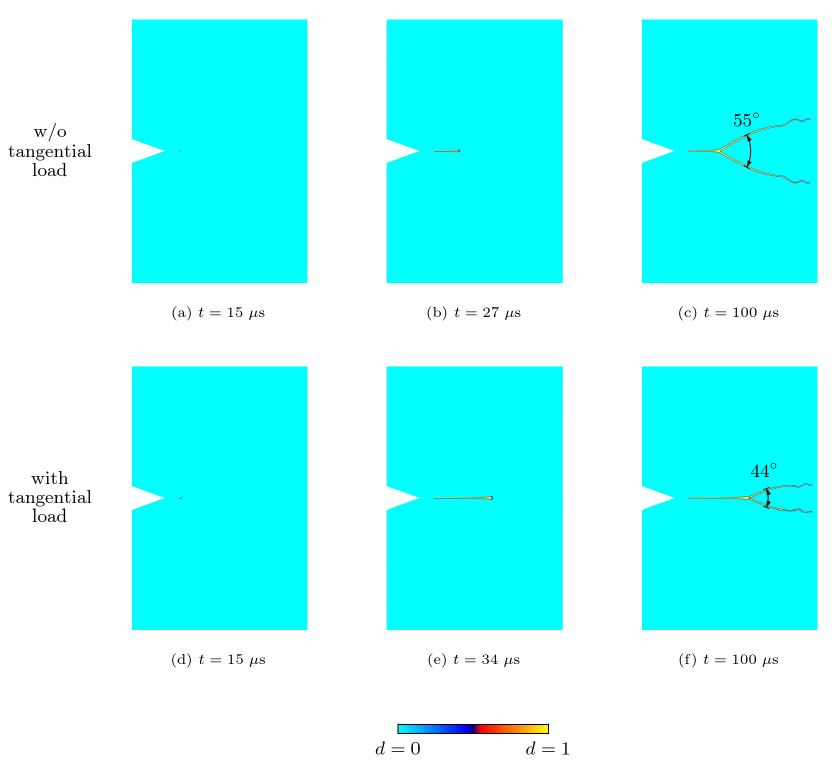}
  \caption{Contour plots of the phase field $d$ in the soda-lime impact tests for cases without tangential load, (a)-(c), and with tangential load, (d)-(f), at selected time steps. The plots are mirrored along the axis of symmetry for visualization purposes.}
  \label{fig:soda_pf}
\end{figure}

\begin{figure}[t!]
  \centering
  \begin{subfigure}[b]{0.45\columnwidth}
    \centering
    \includegraphics[width=0.6\columnwidth]{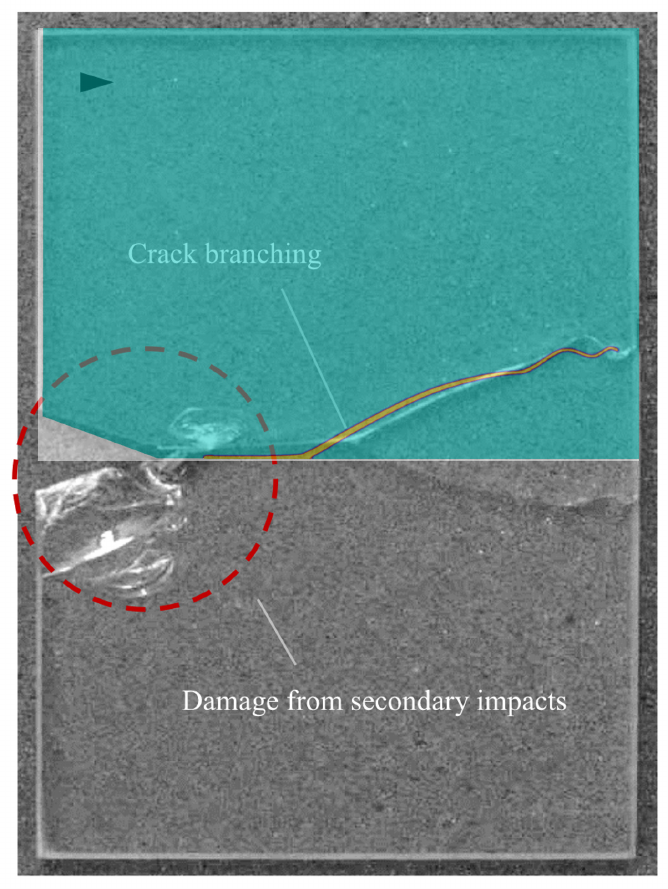}
    \caption{without tangential load}
  \end{subfigure}
  \hfill
  \begin{subfigure}[b]{0.45\columnwidth}
    \centering
    \includegraphics[width=0.6\columnwidth]{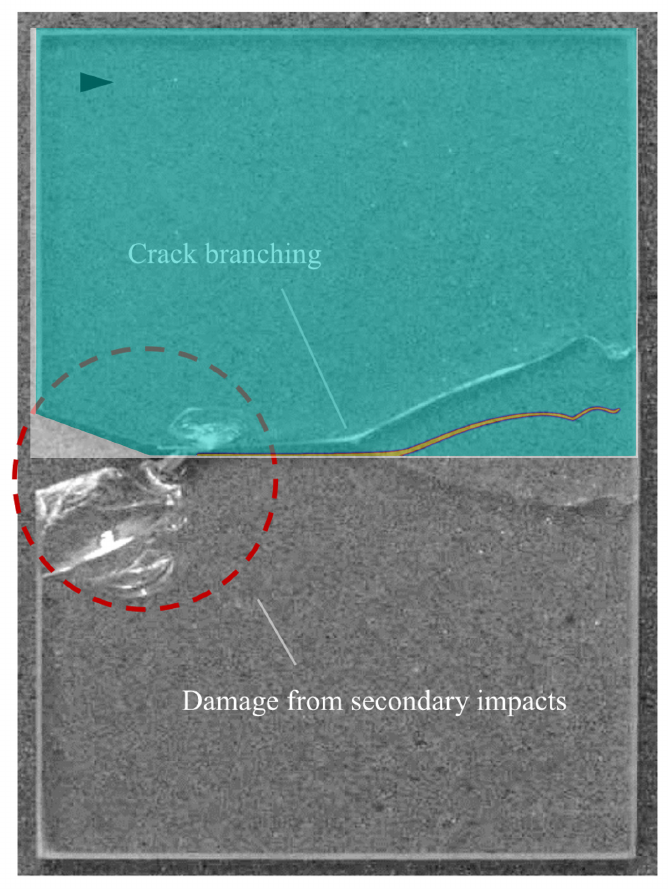}
    \caption{with tangential load}
  \end{subfigure}
  \caption{Comparison of the final phase field contour for (a) the case without tangential load and (b) the case with tangential load with the experimental results from \cite{sundaram_soda_exp_2018}. The phase field contour plots are superimposed on one of the cracked glass specimens from the experiments.}
  \label{fig:soda_pf_vs_exp}
\end{figure}

\begin{figure}[t!]
  \centering
  \includegraphics[width=\columnwidth]{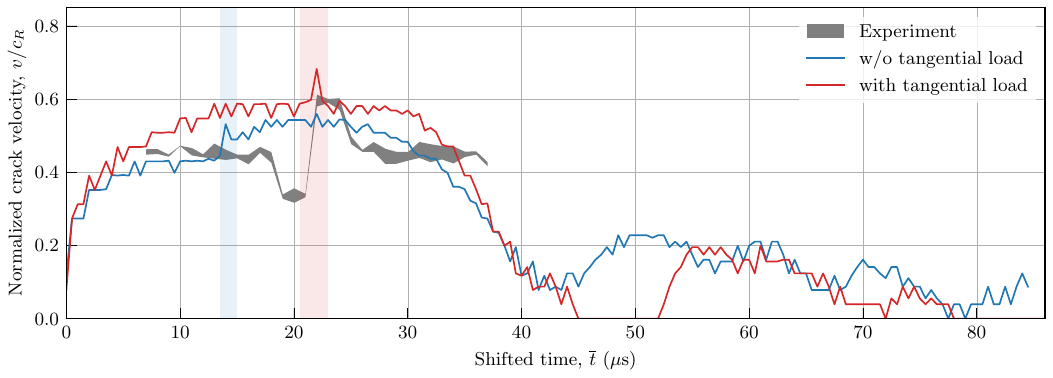}
  \caption{Comparison of crack tip speeds from experiments in \cite{sundaram_soda_exp_2018} and the phase field simulations with and without tangential load. The shifted time $\bar{t}=0$ corresponds to the onset of crack nucleation. The shaded regions (pink and blue) correspond to times in which branching occured in the simulations with and without an applied tangential load on the V-notch faces, respectively.}
  \label{fig:soda_tip_speed}
\end{figure}

Contours of the phase field $d$ from the proposed phase-field model, with and without tangential load, at selected time steps are shown in \figref{fig:soda_pf}. The evolution of the phase field $d$ clearly shows the ability of the model to capture crack nucleation, propagation, and branching of a rapidly evolving dynamic fracture. In addition, the results show that without the tangential load crack branching occurs earlier at $t=27~\mu$s, compared to a time of $t=34~\mu$s with the tangential load. As illustrated in \figref{fig:soda_pf}(c,f), crack branching angles of approximately $54^\circ$ and $44^\circ$ result from the simulations. These angles effectively bracket the experimental observations, in which the  branching angles between $47^\circ$ and $55^\circ$ were reported; see \cite{sundaram_soda_exp_2018,tipper_soda_exp_conference_2020}.

In \figref{fig:soda_pf_vs_exp}, the final crack patterns obtained from the simulations are compared to the experimental results by overlaying the phase field contour with the photograph of the cracked glass specimen. In \figref{fig:soda_pf_vs_exp}, when comparing to the experimental results, we find that for the simulations without the tangential load, crack branching occurs earlier at $(X_1,X_2)=(47, 0)$ with a larger branching angle. In simulations with the tangential load, crack branching occurs later at $(X_1,X_2)=(57, 0)$ with a relatively smaller branching angle. Additionally, small twists near the ends of crack branches are observed in both the simulations and the experiments.


Lastly, we compare the crack tip speeds from the simulations with the experimental observations reported in \cite{sundaram_soda_exp_2018}. The modeled crack speeds exceed experimental values by 10\% to 20\% while remaining below 60\% of the Rayleigh wave speed, except for simulations with the tangential load (and just around the branching time). For simulations with the tangential load, the branching time ($\bar{t}=21 \sim 23~\mu$s) matches the experimental observation well. Furthermore, the modeled post-branching crack speeds rapidly decline around $\bar{t} = 45~\mu$s, followed by fluctuations in crack tip speeds resulting from the superposition reflected stress waves at later times. Those fluctuations in crack tip speeds correspond to the formation of the aforementioned small twists near the ends of the crack branches.


\section*{Acknowledgements}

This work was supported by the National Science Foundation through the Grants  CMMI--2132551 and CMMI--2132528. This support is gratefully acknowledged.

\bibliographystyle{cas-model2-names}
\bibliography{References}

\end{document}